\documentclass[11pt]{article}
\usepackage[a4paper, portrait, margin=2.5cm]{geometry}
\usepackage{graphicx}
\usepackage{lastpage}
\usepackage{fancyhdr}

\fancypagestyle{firstpage}{%
  \setlength\headheight{154pt}
  \fancyhf{}%
  \fancyhead[C]{\includegraphics[width=\textwidth]{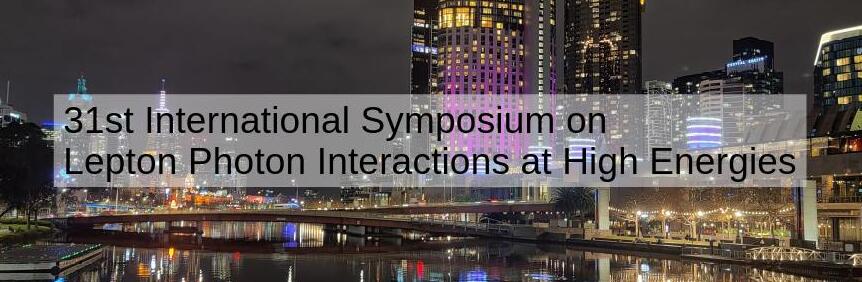}}%
  \fancyfoot[C]{CC-BY-4.0 licence}%
}

\newcommand{\pp}{\pi^+\pi^-}
\newcommand{\ks}{K_S^0}

\newcommand{\EE}{e^+e^-}

\newcommand{\psp}{\psi(2S)}
\newcommand{\psip}{\psi(2S)}
\newcommand{\pspp}{\psi(3770)}
\newcommand{\jpsi}{J/\psi}

\newcommand{\xyz}{\rm XYZ}
\newcommand{\x}{X(3872)}

\newcommand{\zc}{Z_c(3900)}
\newcommand{\zcp}{Z_c(4020)}

\newcommand{\ppjpsi}{\pi^+\pi^-J/\psi}
\newcommand{\hc}{h_c}
\newcommand{\pphc}{\pi^+\pi^-\hc}

\newcommand{\ccb}{c\bar{c}}

\begin{document}
\title{Heavy Flavour Spectroscopy} 
\date{17-21 July 2023}
\author{Chang-Zheng~Yuan \\ Institute of High Energy Physics, Chinese Academy of Sciences, Beijing
\\
University of Chinese Academy of Sciences, Beijing} 

\newgeometry{top=2cm, bottom=7cm}
\maketitle
\thispagestyle{firstpage}

\abstract{
The discovery of hadronic states beyond the conventional two-quark meson
and three-quark baryon picture in the last two decades is one of the 
most amazing accomplishments in fundamental physics research. 
We review the experimental progress on the study of the exotic states
(also known as the $\xyz$ particles) beyond the conventional quark model. 
We give a general review and then focus on the lineshape measurement of 
the $\x$, observation of new decay modes of the $Y(4230)$ and 
new vector charmoniumlike states $Y(4500)$ and $Y(4790)$, 
evidence for the neutral isospin partners of the charged 
charmoniumlike $Z_{cs}$ states, discoveries of the tetraquark 
state candidates with four different flavours or two-pairs of charm-anticharm 
quarks and the pentaquark states.
}
\restoregeometry

\noindent
{\bf\em\boldmath Introduction:}
Hadron spectroscopy is a field of frequent discoveries and surprises,
and the theoretical difficulties in understanding the strong interaction
in the color-confinement regime make the field even more fascinating.
The tremendous data collected by the BaBar, Belle, BESIII, LHCb,
and other experiments and improved theoretical tools developed to analyze the
experimental data result in rapid progress of the field.

In the conventional quark model, mesons are composed of one quark
and one anti-quark, while baryons are composed of three quarks.
However, many quarkoniumlike states were discovered at two
$B$-factories BaBar and Belle~\cite{PBFB} in the first decade of
the 21st century. Whereas some of these
are good candidates of quarkonium states, many other states have
exotic properties, which may indicate that exotic states, such as
multi-quark state, hadronic molecule, or hybrid, have been
observed~\cite{reviews}.

BaBar and Belle experiments finished their data taking in 2008 and
2010, respectively, and the data are still used for various
physics analyses. BESIII~\cite{ijmpa_review} and
LHCb~\cite{lhcb} experiments started data taking
and contributed to the study of exotic hadrons since 2008.
Most of the discoveries of the such states were made at these
four experiments.

\begin{figure*}[htbp]
\centering
  \includegraphics[width=0.90\textwidth]{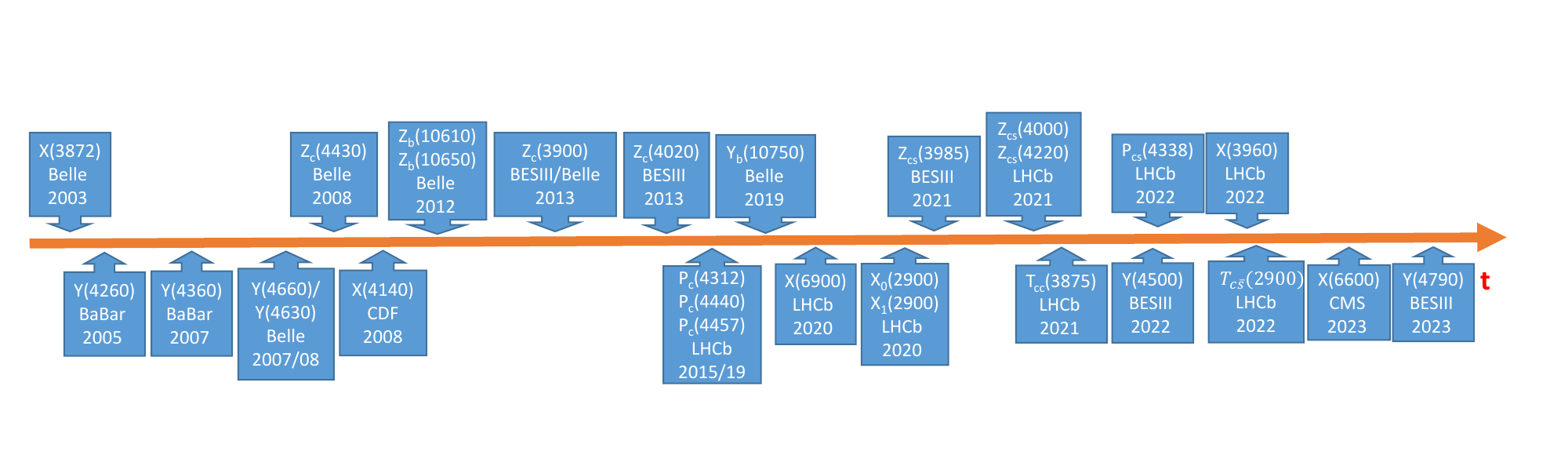}
\caption{Discovery of some heavy exotic states from experiments.}\label{fig:XYZ}
\end{figure*}

Figure~\ref{fig:XYZ} shows the history of the discovery of some
of the new hadrons, started from the observation of the $X(3872)$
in 2003~\cite{Bellex}.
In this article, we present recent experimental progress
and focus on those states with exotic properties, including the
$\x$, $Y(4260)$, $\zc$, $P_c$ and their siblings.

\noindent
{\bf\em\boldmath The lineshape of the $\x$: } 
The $X(3872)$ was observed in 2003 by the Belle experiment~\cite{Bellex}, and 
confirmed very soon by the CDF~\cite{CDFx} and $D0$~\cite{D0x} experiments
in $p\bar{p}$ collision.

The mass of the $X(3872)$ has been measured as $3871.65\pm 0.06$~MeV~\cite{pdg},
which is lower than the mass threshold of $\bar{D}^0D^{*0}$, $3871.69\pm 0.11$~MeV,
by $0.04\pm 0.12$~MeV, to be compared with the bounding energy of the deuteron
of 2.2~MeV. 

The width measurements are less precise and model dependent since
the $X(3872)$ is very narrow and the mass resolution of the experiments
is usually much larger than the intrinsic width. Fitting the
$\ppjpsi$ invariant mass distribution with a Breit-Wigner (BW) function,
LHCb reported a width of about 1~MeV (the mass resolution is 2.4--3.0~MeV);
and the fit with a Flatt\'e function with constraints from other measurements
yields a FWHM of 0.22~MeV which depends strongly on
the $X(3872)\to \bar{D}^0D^{*0}$ coupling~\cite{LHCb_x_width1,LHCb_x_width2}.

\begin{figure*}[hbtp]
\centering
\includegraphics[height=3.5cm]{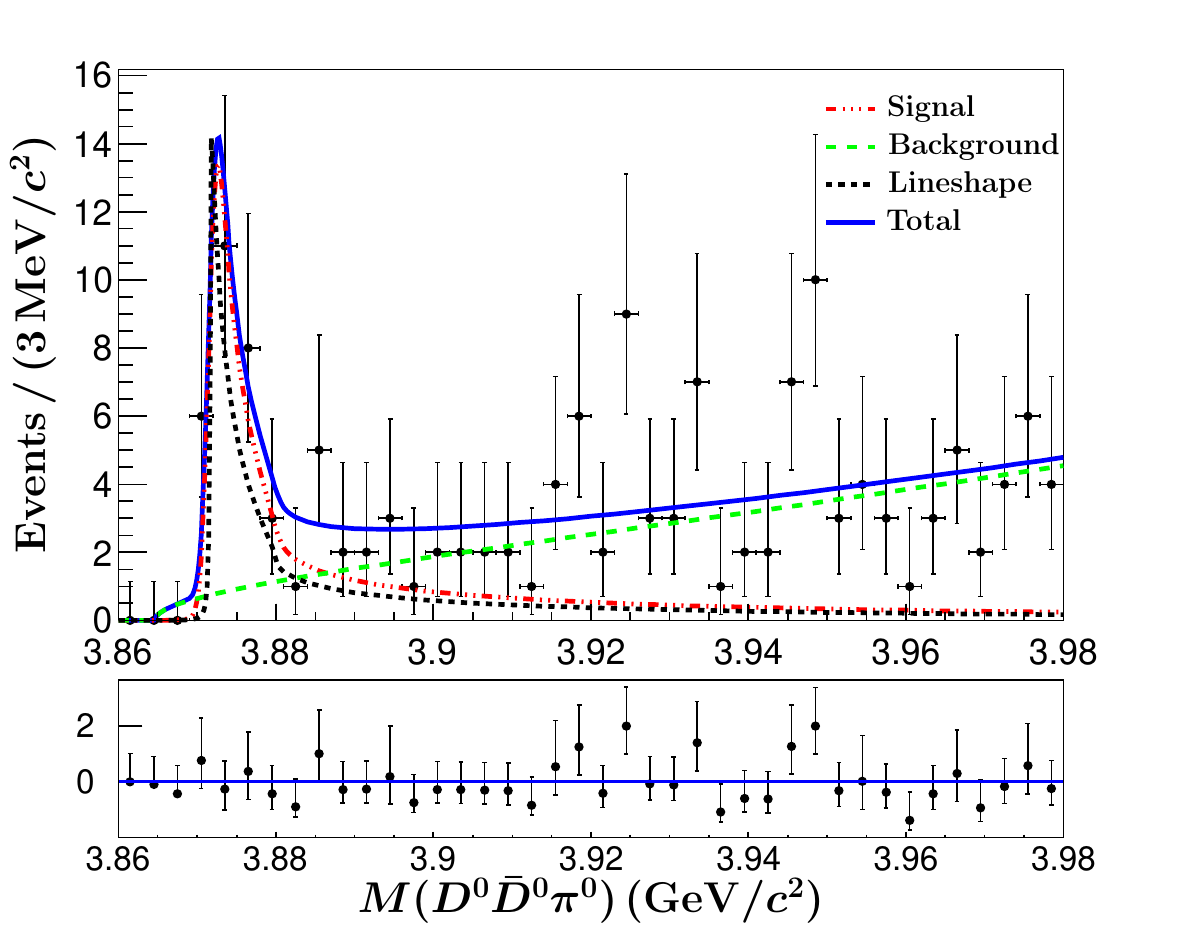}%
\includegraphics[height=3.5cm]{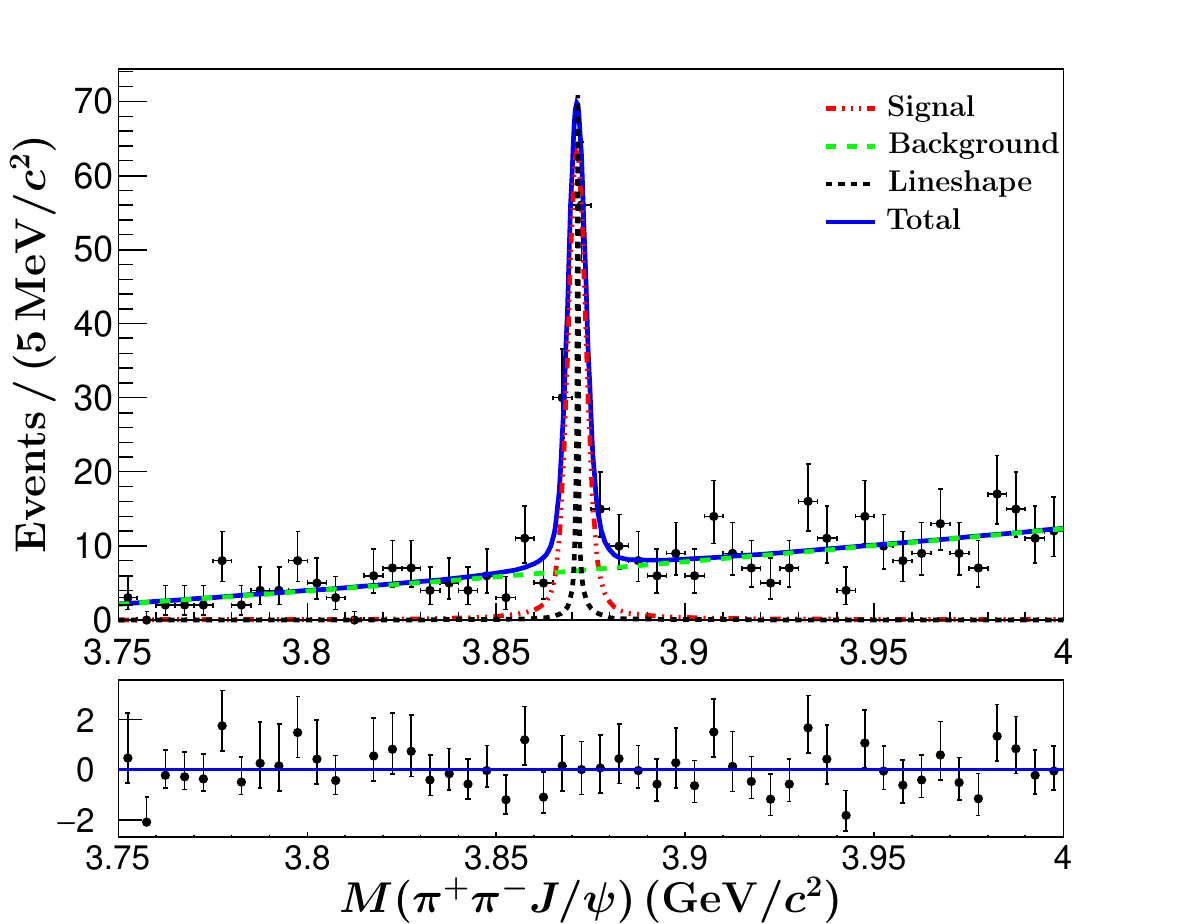}%
\includegraphics[height=3.6cm]{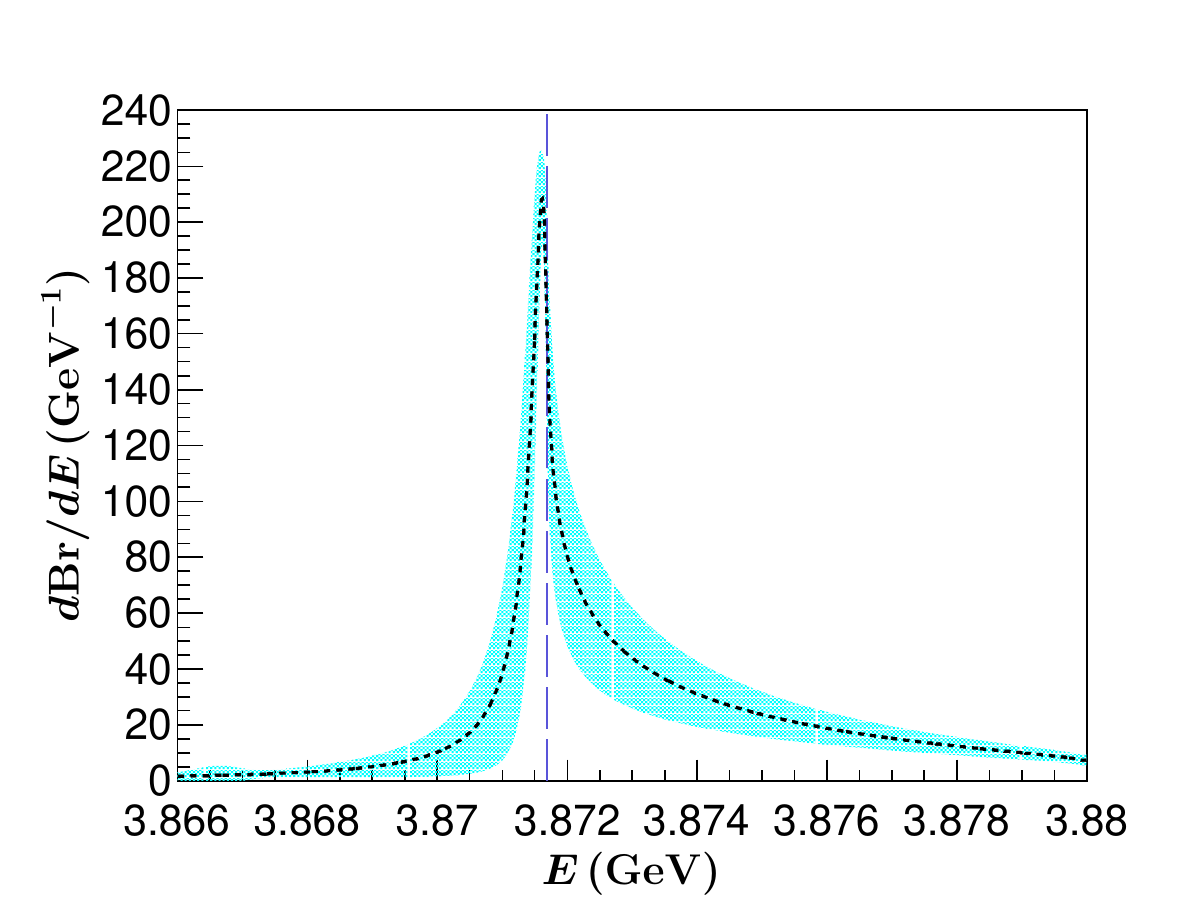}
\caption{\label{fig:x3872} 
The fit to the $D^0\bar{D}^0\pi^0$ (left) and $\pi^+\pi^- J/\psi$ (middle) 
invariant mass distributions. Data are taken from Ref.~\cite{bes3_x_brs}. 
The $X(3872)$ lineshape at the best estimation is shown in right panel. 
The vertical dashed line indicates the position of $D^{*0}\bar{D}^0 $ threshold.}
\end{figure*}

Although the statistics are
low at BESIII experiment, the high efficiencies of reconstructing
all the $X(3872)$ decays modes and the very good mass resolution in
the $\bar{D}^0D^{*0}$ mode ($<1$~MeV) make it possible to measure
the lineshape of the $X(3872)$ state.
BESIII determined the pole locations of the $X(3872)$ based
on a simultaneous fit to the data samples of $X(3872)\to
D^0\bar{D}^0 \pi^0$ and $X(3872)\to\pi^+\pi^- J/\psi$, with the
$X(3872)$ produced in $e^+e^-\to\gamma X(3872)$ process~\cite{bes3_x_brs}.
The parameterization of the $\x$ lineshape, with the effect of
$D^{*0}$ width taken into account, is developed in 
Ref.~\cite{Hanhart:2010wh}. The fit results and the lineshape of 
the $X(3872)$ are shown in Fig.~\ref{fig:x3872}. 
The lineshape parameters are determined to be $g=(0.16\pm0.10^{+1.12}_{-0.11})$,
$\Gamma_0=(2.67\pm1.77^{+8.01}_{-0.82})$~MeV and
$M_{X}=(3871.63\pm 0.13^{+0.06}_{-0.05})$~MeV. 
Here $g$ denotes the effective coupling constant of the $X(3872)$ to neutral and
charged $D^{*}\bar D$; the constant $\Gamma_0$ represents
all the channels except $D^*\bar{D}$, and is separated into three
parts: $\Gamma_0=\Gamma_{\pi^+\pi^- {J}\slash \psi}+\Gamma_{\rm
known}+\Gamma_{\rm unknown}$; and $M_X$ is the mass of the $X(3872)$.
The FWHM of the lineshape is determined to be
$(0.44^{+0.13~+0.38}_{-0.35~-0.25})$~MeV.
Two poles are found on the first and second Riemann sheets corresponding 
to the $D^{*0}\bar{D}^0$ branch cut. The pole location on the first sheet
is much closer to the $D^{*0}\bar{D}^0$ threshold than the other,
and is determined to be $(7.04\pm0.15^{+0.07}_{-0.08})$~MeV above
the $D^0\bar{D}^0\pi^0$ threshold with an imaginary part
$(-0.19\pm0.08^{+0.14}_{-0.19})$~MeV.

Belle measured the $X(3872)$ lineshape with $B \to
X(3872)K\to D^0\bar{D}^{*0}K$~\cite{Belle:2023zxm}. 
The peak near the threshold in the $D^0\bar{D}^{*0}$ 
invariant mass spectrum is fitted using a relativistic 
BW function. Belle determined a
mass of $(3873.71 ^{+0.56}_{-0.50}\pm0.13)$~MeV and a
width of $(5.2^{+2.2}_{-1.5}\pm 0.4)$~MeV. 
The peak is also studied using a Flatt\'{e} lineshape and 
the lower limit on the $D\bar{D}^{*}$ coupling constant 
$g$ is determined to be $0.075$ at 95\% credibility. 
A coupled channel analysis of the data used in this analysis and 
those in $\x\to \ppjpsi$ decay is highly recommended to get reliable 
information about the $\x$ lineshape.

\noindent
{\bf\em\boldmath Observation of new $\psi(4230)$ decays and new vector states $Y(4500)$ and $Y(4790)$: } 
The $Y$ states were discovered in the initial state radiation in
the $B$-factory experiments, and they have $J^{PC}=1^{--}$.
So these state can also be produced directly in $\EE$ annihilation
experiment like BESIII. Much improved measurements of the 
$Y(4260)$~\cite{babar_y4260}, $Y(4360)$, $Y(4660)$~\cite{belle_y4660} 
and so on are achieved, their new decay modes are discovered and new 
vector states are observed.

The most precise measurements of the $Y(4260)$ are from the BESIII
experiment~\cite{BESIII:2016bnd,BESIII:2022qal}. By doing a high
luminosity energy scan in the vicinity of the $Y(4260)$,
BESIII found the peak of the $Y(4260)$ is much lower (so it is now 
named the $\psi(4230)$) than that from
previous measurements and the width is narrow, and there is a high
mass shoulder with a mass of 4.32~GeV if fitted with a BW
function. Since then, more new decay modes of the $\psi(4230)$ were
observed including $\pphc$\cite{BESIII:2016adj}, $\pp\psp$~\cite{BESIII:2021njb}, 
$\omega\chi_{c0}$~\cite{BESIII:2019gjc}, $\pi \bar{D} D^*+c.c.$~\cite{BESIII:2018iea}, 
$\pi \bar{D}^* D^*$~\cite{BESIII:2023cmv}, and 
$K\bar{K}\jpsi$~\cite{BESIII:2022joj,BESIII:2022kcv}.

The cross sections of $e^+e^-\rightarrow K^+K^-J/\psi$ at
center-of-mass energies from 4.127 to 4.600~GeV are measured based
on 15.6~fb$^{-1}$ data collected by the BESIII experiment~\cite{BESIII:2022joj}. 
Two resonant structures are
observed in the line shape of the cross sections. The mass and
width of the first structure are measured to be
$4225.3\pm2.3\pm21.5$~MeV and ($72.9\pm6.1\pm30.8$)~MeV,
respectively. They are consistent with those of the established
$\psi(4230)$. The second structure is observed for the first time
with a statistical significance greater than 8$\sigma$, denoted as
$Y(4500)$. Its mass and width are determined to be
$4484.7\pm13.3\pm24.1$~MeV and $111.1\pm30.1\pm15.2$~MeV,
respectively. This state is confirmed in $\EE\to \pi \bar{D}^* D^*$ 
reported in Ref.~\cite{BESIII:2023cmv}.

With the world's largest $e^+e^-$ scan data sample
between $4.226$ and $4.95$~GeV accumulated by BESIII, the Born
cross sections of $e^+e^-\to D_s^{\ast+}D_s^{\ast-}$ are 
measured precisely~\cite{BESIII:2023wsc}. 
Besides two enhancements in the energy dependent cross sections at 
around 4.2 and 4.45~GeV that may come from the $\psi(4160)$ or $\psi(4230)$ 
and the $\psi(4415)$, respectively, a third resonance structure ($Y(4790)$) 
is observed at around 4.7$\sim$4.8~GeV with statistical significance 
greater than $6.1\sigma$. Due to the limited number of data
points around 4.79~GeV, the fitted mass of the third structure
varies from 4786 to 4793~MeV and the width from 27 to
60~MeV. This could be the same state observed in $\EE\to \ks\ks\jpsi$
with a statistical significance of $4.0\sigma$~\cite{BESIII:2022kcv}.

In the charmonium energy region between 3 and 5~GeV, we now have identified
6 well known $\psi$ peaks ($\jpsi$, $\psip$, $\pspp$, $\psi(4040)$, $\psi(4160)$, 
and $\psi(4415)$) and 9 new $Y$ structures ($Y(4230)$, $Y(4320)$, 
$Y(4360)$, $Y(4390)$, $Y(4500)$, $Y(4630)$, $Y(4660)$, 
$Y(4710)$, and $Y(4790)$). They are all vector states and they cannot be 
all charmonium states. While more experimental efforts are needed to
resolve the origins of these states, theoretical efforts are also
necessary to identify if the vector charmonium hybrids and/or tetraquark states
have already been observed.

\noindent
{\bf\em\boldmath Neutral partners of charged charmoniumlike $Z_{cs}$ states: } 
The charged charmoniumlike state $Z_c(3900)$ discovered in $\pi\jpsi$ by 
the BESIII~\cite{zc3900} and Belle~\cite{Belle_zc} experiments, the $Z_c(4020)$ 
discovered in $\pi\hc$ by the BESIII~\cite{BESIII:2013ouc} experiment,
and the $Z_c(4430)$ discovered in $\pi\psp$ by the Belle~\cite{Belle_zc4430}
experiment are all states with minimal quark content of $\ccb u\bar{d}$. 

Recent studies try to search for states with one of the four quarks 
replaced by a different quark, for example, the $Z_{cs}$ states 
with quark content $\ccb u \bar{s}$.
BESIII announced observation of a
near-threshold structure $Z_{cs}(3985)$ in the $K^+$ recoil-mass spectra
in $\EE\to K^+(D^-_sD^{*0}+D^{*-}_sD^{0})$~\cite{BESIII:2020qkh} with a mass of
3983~MeV and a width of about 10~MeV; and LHCb reported two
resonances decaying into $K^\pm\jpsi$, the $Z_{cs}(4000)$ with a mass
of 4003~MeV and a width of about 131~MeV, and the $Z_{cs}(4220)$
with a mass of 4216~MeV and a width of about 233~MeV~\cite{LHCb:2021uow}.
The widths of the $Z_{cs}(3985)$ and $Z_{cs}(4000)$ 
are quite different, maybe one of them is the strange partner of the 
$Z_c(3900)$ with the $d$ quark replaced with an $s$ quark.
Both BESIII~\cite{BESIII:2022qzr} and LHCb~\cite{LHCb:2023dhg} reported 
evidence for the neutral partners of the $Z_{cs}$ states at around 4~GeV
with quark content $\ccb u \bar{s}$. These indicate that these states 
form isospin doublets. 

\begin{figure*}[htbp]
\centering
  \includegraphics[width=0.55\textwidth]{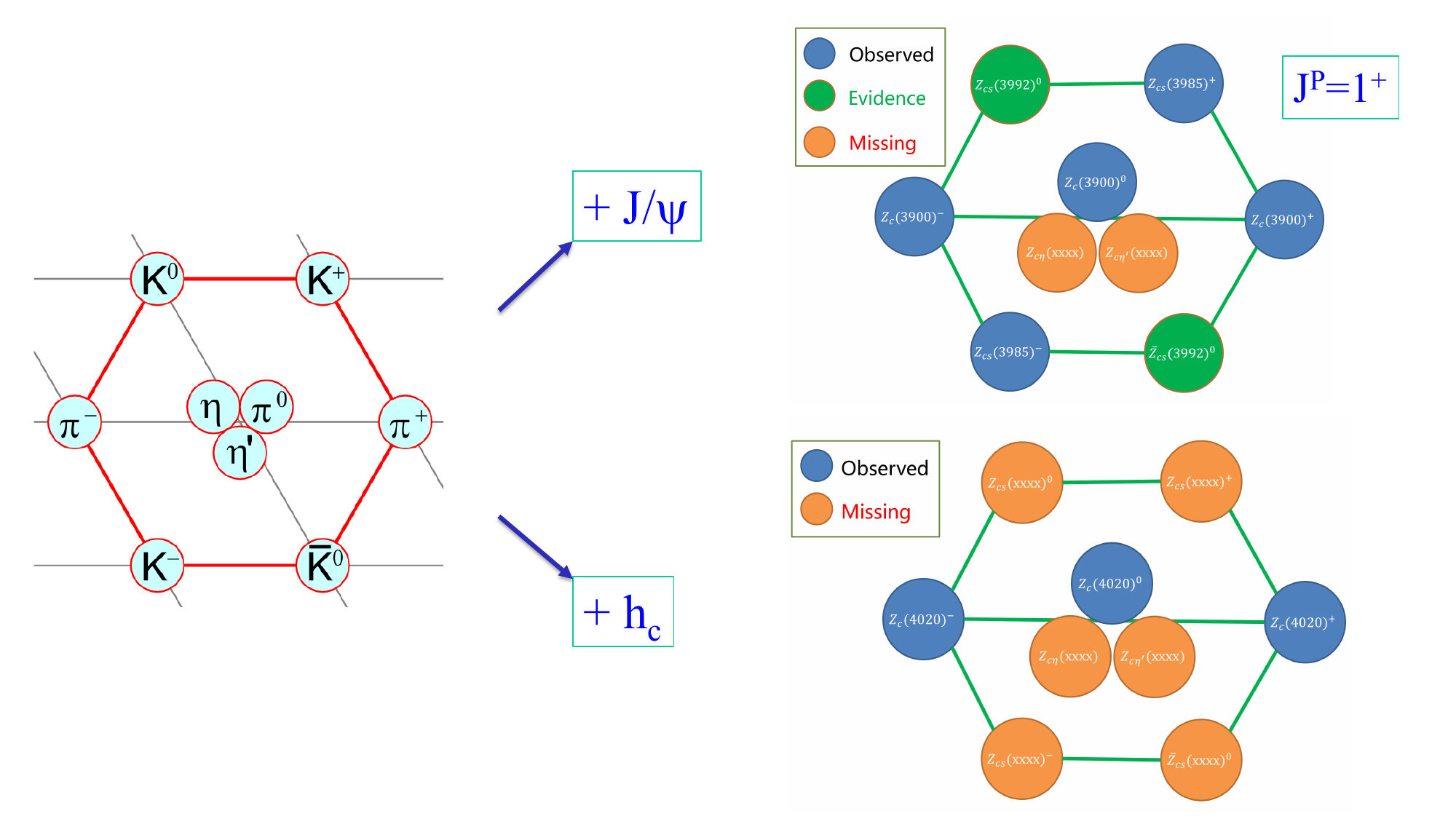}
\caption{The possible multiplets of the $Z_c(3900)$ and $Z_c(4020)$.}\label{fig:zc}
\end{figure*}

The $Z_c(3900)$ ($Z_c(4020)$) and $Z_{cs}$ states may form multiplets 
shown in Fig.~\ref{fig:zc}, the missing states can be searched for 
with the existing or future data samples.

\noindent
{\bf\em\boldmath New tetraquark states from LHC experiments: }
LHCb experiment observed two new resonances with four different flavors
with mass of $2908\pm11\pm20\,\rm{MeV}$ and width of $136\pm23\pm13\,\rm{MeV}$, 
which decay to $D^+_s\pi^+$ and $D^+_s\pi^-$, respectively,
from a combined amplitude analysis for the decays 
$B^0 \to \overline{D}{}^0 D^+_s\pi^-$ and 
$B^+\to D^- D^+_s\pi^+$, which are related by isospin symmetry. 
The former state indicates the first observation of a doubly charged 
open-charm tetraquark state with minimal quark content $c\bar{s}u\bar{d}$, 
and the latter state is a neutral tetraquark composed of $c\bar{s}\bar{u}d$ 
quarks ($T_{c\bar{s}}$). Both states are found to have spin-parity $0^+$,
and their resonant parameters are consistent with each other, 
which suggests that they belong to an isospin triplet~\cite{LHCb:2022sfr}.

Tetraquark states $T_{cs}$ with four different flavors ($cs\bar{u}\bar{d}$) 
have been search for at LHCb and evidence ($3.9\sigma$) for two states 
($X_0(2900)$ and $X_1(2900)$) in $D^-K^+$ system were reported from a PWA
of $B^+\to D^+D^-K^+$ events by the LHCb experiment~\cite{LHCb:2020bls}. 
They are good candidates for the flavour partners of the $T_{c\bar{s}}$ states, 
and more flavour partners with other quark contents and spin-parities 
are expected.

The LHCb, ATLAS, and CMS experiments reported observation of states 
decay to two charmonium states~\cite{LHCb:2020bwg,ATLAS:2023bft,CMS:2023owd}.
The $X(6900)$ is observed in all these three experiments, and 
a new structure ($X(6600)$), with a significance above $5\sigma$, and
evidence for another new structure ($X(7300)$), with a local significance of
$4.1\sigma$, are found at CMS. The masses, widths, and significances are 
obtained in model-dependent ways without considering possible interference 
between the resonances. These are good candidates for tetraquark states 
with two pairs of charm-anticharm quarks.

\noindent
{\bf\em\boldmath Pentaquark states: }
In the decay of $\Lambda_b \to \jpsi p K^-$ analyzed by the LHCb experiments,
there are three very narrow peaks in the invariant mass distribution of
$\jpsi p$~\cite{LHCb:2019kea}. In a simple fit to the invariant mass spectrum, 
the resonance parameters of the $P_c(4312)^+$, $P_c(4440)^+$, and $P_c(4457)^+$ are 
determined. They are all narrow, and the $P_c(4312)^+$ state peaks right 
below the $\Sigma_c^+\bar{D}^0$ threshold, the $P_c(4457)^+$ state 
peaks right below the $\Sigma_c^+\bar{D}^{*0}$ threshold, 
while the $P_c(4440)^+$ state peaks about 20~MeV below it.

Being so close to the thresholds, they are very good candidates 
for the molecules of a charmed baryon and an anti-charmed meson,
and more similar states close to the other baryon-meson 
thresholds are expected. In a following amplitude analysis 
of $B \to \jpsi \Lambda \bar{p}$ decays, a narrow resonance 
in the $\jpsi \Lambda$ system, consistent with a pentaquark 
candidate with strangeness, is observed with high significance~\cite{LHCb:2022ogu}. 
The mass and the width of this new state are measured to be 
$4338.2\pm 0.7\pm 0.4$~MeV and $7.0\pm 1.2\pm 1.3$~MeV, respectively.
It is very close to the $\Xi_c^+ D^-$ threshold of 4337.4~MeV.
Evidence for states at around $\Xi_c^0 \bar{D}^{*0}$ threshold 
are reported in Ref.~\cite{LHCb:2020jpq}.

\noindent
{\bf\em\boldmath Summary and Perspectives: }
Many states with exotic properties were observed in the past two decades.
Some of them are quite close to the thresholds of two heavy objects, 
either two heavy flavor mesons or one heavy flavor meson and one heavy 
flavor baryon, like the $X(3872)$ ($\bar{D}^0D^{*0}$),
$Y(4220)$ ($D_s^{*+}D_s^{*-}$ or $\bar{D}D_1)$,
$\zc^+$ ($\bar{D}^0D^{*+}$),
$\zcp^+$ ($\bar{D}^{*0}D^{*+}$),
$Z_{cs}^+$ ($\bar{D}^0D_s^{*+}$),
$P_c(4312)^+$ ($\Sigma_c^+\bar{D}^0$), 
$P_c(4440)^+$ and $P_c(4457)^+$ ($\Sigma_c^+\bar{D}^{*0}$); 
and some other states are not close to such thresholds, such as 
the $Y(4360)$, $Y(4660)$, $Z_c(4430)^+$, and $Z_{cs}(4220)^+$.
These may suggest that we did observe the hadronic molecules
close to thresholds and we also observed hadronic states
with some other quark configurations like compact tetraquark
states and so on. It is expected that more results will be
produced by the Belle II, BESIII, LHCb, and other experiments.

\noindent
{\bf\em\boldmath Acknowledgments: }
We thank the organizers for the invitation to give a talk at
the conference in such a beautiful city. This work is supported in part
by National Key Research and Development Program of China
(No.~2020YFA0406300), and National Natural Science Foundation
of China (NSFC, Nos. 11961141012 and 11835012).

\end{document}